# Segmentation of Lungs COVID Infected Regions by Attention Mechanism and Synthetic Data

Ali Zindari[1*], Parham Yazdekhasty[1*], Zahra Nabizadeh-ShahreBabak[1], Pejman Khadivi[2], Nader Karimi[1], Shadrokh Samavi[1,3]

[1]*Department of Electrical and Computer Engineering, Isfahan University of Technology,* Isfahan, Iran
[2]*Computer Science Department, Seattle University,* Seattle, USA
[3]*Department of Electrical and Computer Engineering*, *McMaster University,* Hamilton, Canada

*Abstract* - **Coronavirus has caused hundreds of thousands of deaths. Fatalities could decrease if every patient could get suitable treatment by the healthcare system. Machine learning, especially computer vision methods based on deep learning, can help healthcare professionals diagnose and treat COVID-19 infected cases more efficiently. Hence, infected patients can get better service from the healthcare system and decrease the number of deaths caused by the coronavirus. This research proposes a method for segmenting infected lung regions in a CT image. For this purpose, a convolutional neural network with an attention mechanism is used to detect infected areas with complex patterns. Attention blocks improve the segmentation accuracy by focusing on informative parts of the image. Furthermore, a generative adversarial network generates synthetic images for data augmentation and expansion of small available datasets. Experimental results show the superiority of the proposed method compared to some existing procedures.**

## 1. Introduction

In December 2019, the first human case of coronavirus was reported in Wuhan. Since then, the world has been dealing with the challenges caused by the coronavirus pandemic. One of the biggest challenges for authorities is decreasing the burden on the healthcare system. A convenient way of doing this is to use machine learning models to detect infected individuals. Since the start of the pandemic, many methods have been proposed to solve this problem. Although the dominant approach has been convolutional neural networks (CNNs) some techniques used classic methods.

In [1], Oulefki et al. proposed a classic approach in contrast to deep learning algorithms. Some pre-processing algorithms are used for image enhancement at the first step, then a multi-stage threshold is applied based on Kapur entropy and combined these thresholds to generate the final mask. In [2], a modified U-net architecture is proposed for segmenting the infected parts of CT images. In the encoder of this architecture, two new blocks are added called Feature Variation (FV) and Progressive Atrous Spatial Pyramid Pooling (PASPP) blocks. The FV block has a trifurcation architecture that extracts PASPP block features that are responsible for extracting some semantic information and follows a residual architecture. In [3], Wang et al. proposed a framework for COVID-19 lesion segmentation with different sizes. They introduced a network which they called COPLENet. An important feature of this method is its noise robustness. Therefore, they designed a new loss function for this task that is resistant to noise. In [4], Wu et al. proposed a model capable of doing both classification and segmentation. The segmentation part is following U-net architecture. The encoder uses a VGG backbone with some modifications. A module called Enhanced Feature Module (EFM) was used in the encoder. They claimed that using this module leads to a better representational power. In [5], Qiu et al. proposed a lightweight and accurate network and can be trained quickly. They also proposed Attentive Hierarchical Spatial Pyramid (AHSP) module and leveraged multi-scale learning as well.

In [6], Zheng et al. incorporated spatial channel attention mechanisms, Pyramid convolution block, and the idea of adding shortcut connections embedded within the U-net. It helped to increase the performance of the simple U-net architecture. Yao et al. in [7] introduced a label-free method base on anomaly detection. They designed a technique to generate lesion-like appearances. At first, lesion-like shapes are generated, and then a lesion-like texture is added to it. Adding these fake lesions to healthy lung CT images made paired training datasets train their model. In [8], Saeedizadeh et al. developed their method based on transfer learning due to the shortage of data for this task. At the main part of their method, U-net architecture is used. Total variation (TV), as an additional term, was added to the loss function to generate more continuous segmented masks. The method presented in [9] is designed for COVID-19 detection (a classification task). However, it also uses segmentation techniques to extract the region of interest (ROI). Then ROI helps to predict better and more precise outputs in the final stages. In [10] authors proposed a deep model named COVID-Net with high architectural diversity. They examined it on two open datasets by merging them and training the model on it.

In this paper, we propose a new architecture to detect infected lung regions in CT images accurately. The proposed method is an improved version of the architecture that we presented in [11]. The changes in our current method are twofold. First, we are adding attention blocks to the network.

---
[*] Ali Zindari and Parham Yazdekhasty have equal contributions

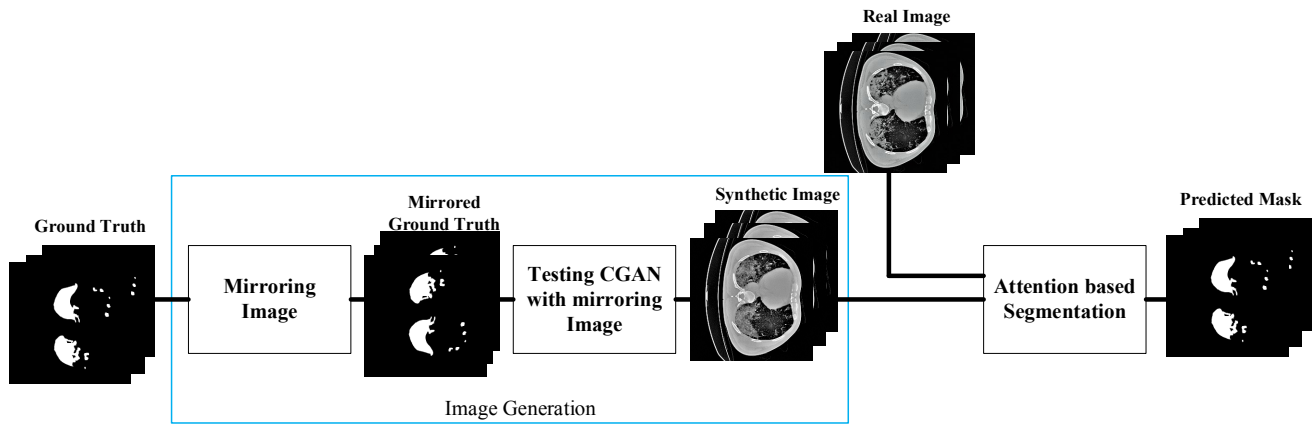

Figure 1: Overview of our method

Moreover, we are using a generative adversarial network for data augmentation. The overview of our work is illustrated in Figure 1.

The rest of the paper is organized in the following manner. In Section 2, we present our data augmentation approach. Section 3 explains the proposed segmentation method. Experimental results are presented in Section 4. Section 5 is dedicated to concluding remarks.

## 2. Proposed Data Augmentation

Image generation is one of the problems that GANs are one of the effective solutions for it. By using this technique, the image of the dataset could be increased. Due to this, these models could be used for data augmentation. Data augmentation is a technique that is used for increasing the amount of data while we are suffering from a lack of data. The Covid-19 datasets that are publically available are almost having this problem. Since large datasets are needed for learning more accurate and robust deep learning models, image generation is used for data augmentation in our model. Many data augmentation techniques are proposed and widely used in many problems related to image processing like image classification or image segmentation. Generally, we can classify these techniques into two categories: classical approach and deep learning approach.

In classical approaches for data augmentation, we can perform some geometric transformations like flipping, rotating, zooming in/out, or performing some illumination-based changes [12] to increase data.

In deep learning methods, one of the best approaches for data augmentation is GANs. GAN was first introduced by Goodfellow [13] for generating realistic fake images from a random noise input. Then, conditional GAN (cGAN) was proposed in [14], creating data with a condition on the input. By growing up the applications of GANs, new methods for an image-to-image translation are proposed. For example, in [15], authors proposed a translation architecture for paired data, and in [16], authors proposed a model for unpaired data. In [15], the authors proposed a cGAN architecture, which converted an image from one domain to another using paired data. By presenting these methods, many researchers used them to generate synthetic data by giving a synthetic ground truth to the model and generating realistic images from that [17,18].

In this paper, we use both classical and deep learning methods for data augmentation. Because of having a paired dataset, we use the method of [15] for augmentation. Cycle GAN could also be used for this task as well, but it is computationally expensive, and it is better for the datasets that have unpaired data; hence we decide to use conditional GAN (cGAN). Besides this, we use a classical approach method for mirroring the original data. For training a cGAN for data augmentation, we need to make a paired dataset of two domains such as A and B, and then feed them to the cGAN architecture to perform image to image translation (for example, converting an image from domain A to B). As our first domain, let's say A, we used the infected region of each slice by multiplying the CT image to its corresponding binary ground truth. The second domain, B, is the binary ground truth. We decided to generate only the infected parts of the image instead of the whole CT image because the infected parts are of importance. Hence, we have our paired dataset. After that, we feed them to the cGAN architecture proposed in [15] to learn the process of converting a binary image to a realistic synthetic Covid-19 infected region. Therefore, we are converting images from domain B to A. This process is shown in Figure 2.

After the training process, the network has learned the conversion process. Then a binary mask could be fed to the network, and its corresponding infected-region map is generated. However, if we feed the exact slice that the network is trained by it, we will be given the output, which is usually the same as the one in the dataset. Therefore, to generate synthetic images, we mirror each binary mask first then fed it into the network. This process creates some new data that can be added to the dataset. Details of generating synthetic images are shown in Figure 3. In this process, we use the classical mirroring approach for augmentation. The process consists of mirroring a real image and its ground truth. From the mirrored ground truth map, a map of infected regions is generated. Finally, infected pixels are placed in the mirrored real image to generate a synthetic image.

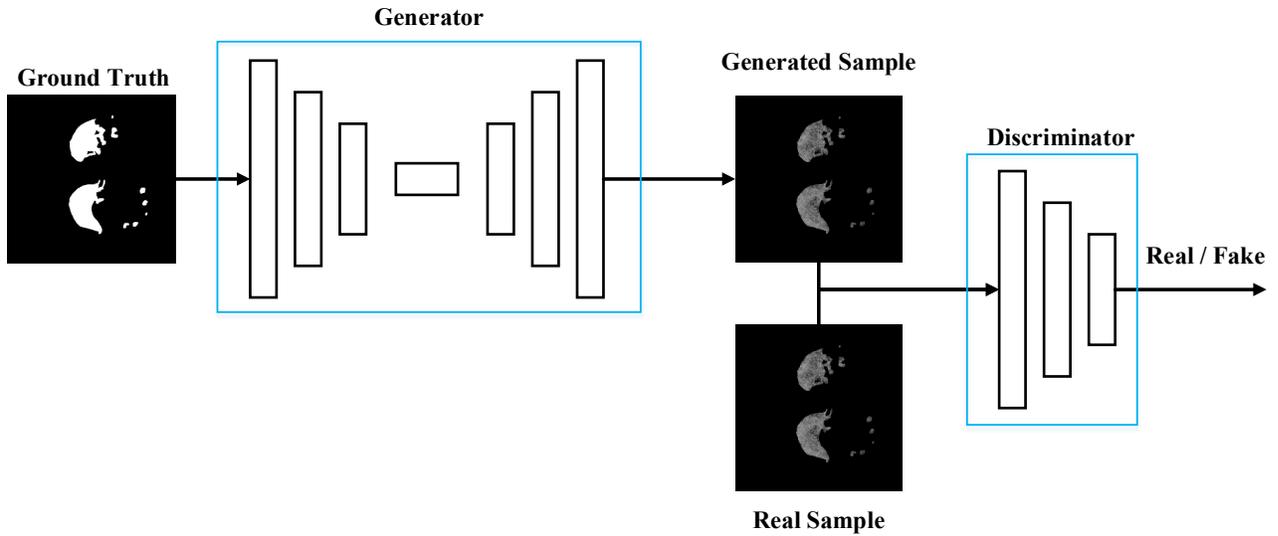

Figure 2. Conditional GAN for infected regions generation

## 3. Image Segmentation

Our proposed method is the enhanced version of the proposed method in [11]. The proposed architecture in [11] is a combination of U-net [19] architecture and multi-task learning [20] concept. The skip connections, which transfer the low-level information in the U-net, make it suitable for segmentation.

All infected regions are located inside the lung region of the CT scan image. However, there is no guarantee that the network only searches inside the lung region. Hence, we need to guide the network to search for infected parts inside the lungs. We will use multi-task learning to force the network to explore specific areas of the image. The network will generate two outputs: first, a mask is generated that shows all possible infected regions of the image and a second mask that shows the lung regions of the CT image. It also uses a modified version of Inception Block inspired from [21].

### 3.1. Attention mechanism

In the field of image processing with deep learning methods, the key component is a convolutional layer. Each pixel of a convolutional layer's output is dependent on the values of a specific neighborhood of pixels in its input feature maps. In contrast, a fully connected layer uses all the input image information to generate its output. Therefore, a convolutional layer needs fewer parameters to be learned compared to a fully connected layer. However, although using the convolutional layer will let us design and train networks with deeper and more complex structures, the convolutional layer usually ignores the contextual information of its input feature maps [22]. A common approach to solve this problem is reducing the image size, layer by layer, using pooling to increase the dependency of each output pixel to the larger neighborhood of input image pixels.

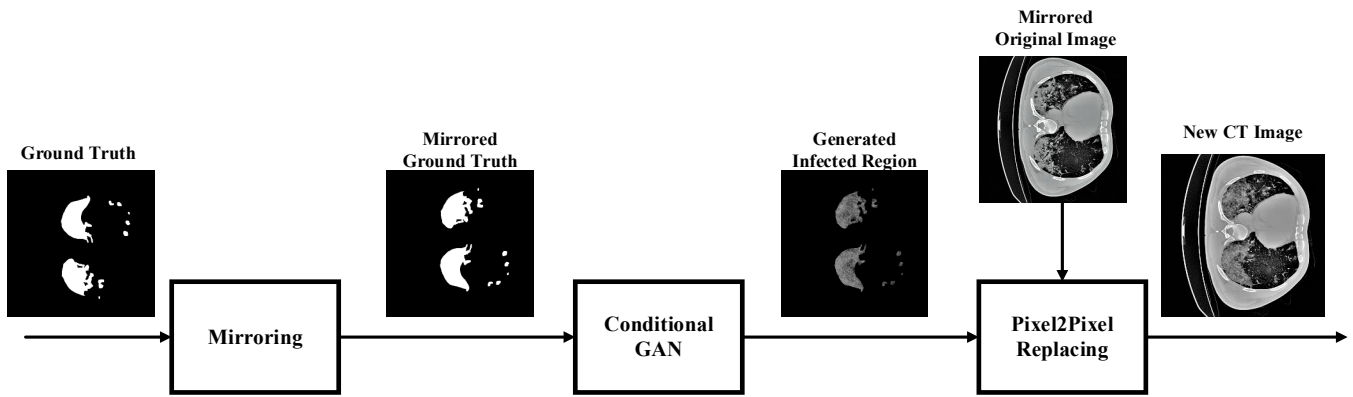

*Figure 3:* Mirrored ground truth map as the input of a Conditional GAN for generation of synthetic images with infected regions.

For the task of covid-19 infected regions segmentation, both local and contextual information is needed. Hence, employing a method that gives the network the ability to extract contextual information can help to improve accuracy. The attention mechanism is another technique that can help the deep network extract contextual information besides the local information. Attention mechanism can be seen as a module whose duty is to summarize all input information and reduce it to a reasonable size. Then the summarized information somehow affects the main information flow in the network. Moreover, using an attention mechanism can be justified if we look at it from a different angle. In feature maps generated in each layer of the network, all pixels do not have the same importance. The same statement is true for channels of feature maps. For example, in our task, brighter regions inside the lungs' CT images probably have more valuable information for the network than dark areas surrounding the lungs. Hence, the attention mechanism can be implemented in two ways. First, to extract the importance of channels of a feature map. Second, the purpose would be to infer the importance of each pixel of a feature map in all channels.

Channel-wise attention is introduced in [23]. We modified it to increase its learning ability by increasing its learnable parameters. Suppose X is a tensor with n channels and the input of the channel-wise attention mechanism. The Channel-wise attention mechanism summarizes information of each channel into two vectors. The first vector is the maximum, and the other one is the mean of the channel's pixels. Hence, we have a 2*n matrix, which represents all the information in X. In the next step, we consider the 2*n matrix as the input of a small fully convolutional network that uses 2*1 filters to summarize it to a 1*n vector. In the final step, a reasonably small, fully connected network learns to convert the previous step output vector to a new vector named W. Each element of W should represent the importance of each channel. In other words, the higher value of $W_i$, the more critical its channel is. Each X channel is multiplied by its corresponding element in W to produce the final output. As a result, channels with smaller $W_i$ will fade away, and channels with larger $W_i$ will be strengthened. In other words, the network can emphasize some more important channels of X. The architectural details of our channel-wise attention mechanism can be seen in Figure 4. This mechanism is the modified version of the proposed mechanism in [23]. The modified blocks are shown in blue color.

Although the base idea of spatial attention mechanism is the same as the channel-wise attention mechanism, they have different approaches to implement it. In contrast to channel-wise attention, which summarizes each channel of its input to two numbers, spatial attention does not destroy the spatial information in each channel by doing that. However, we need to reduce the size to extract contextual information. Spatial attention mechanisms do this by summarizing its input X (a tensor with n channels) to X', which is a tensor with one channel. Like channel-wise attention, it uses maximum and mean operations to do this. We use the same architecture, which is presented in [23].

### 3.2. Attention Fusion

In two previous sections, we explained two attention modules that we have used in our architecture. In the first step, input feature maps go through the channel-wise attention module. After that, the result is multiplied by input feature maps, which assign a specific value to each channel. Next, the output goes through the spatial attention module. The result is again multiplied by the output of the previous part. Hence, we assign specific weight to each pixel in feature maps.

At last, there is a convolution layer that is applied to the output of the previous part. Finally, the outcome is produced by summing the result of this part and the input feature maps. The block diagram of the attention fusion process is shown in Figure 5. This process is different compared to the work of [23]. The fuse attention module is used only in the encoder part of the main network's structure. It is embedded in each level of the encoder after the convolutional layers and before the max pooling. The intuition for using this module just in the encoder part is keeping the model complexity as low as possible to avoid overfitting. Moreover, the encoder part is shared. Hence, improving the encoder leads to improvement in both lung segmentation and the infected region segmentation.

### 3.3. Post Processing

Usually, in lung CT images, some textures are very similar to infected parts, which can be mistaken for COVID-19 infected regions. There is a 98% probability that the area of infected regions is greater than 30 pixels. Hence, we consider any prediction with an area of fewer than 30 pixels as a non-infected slice.

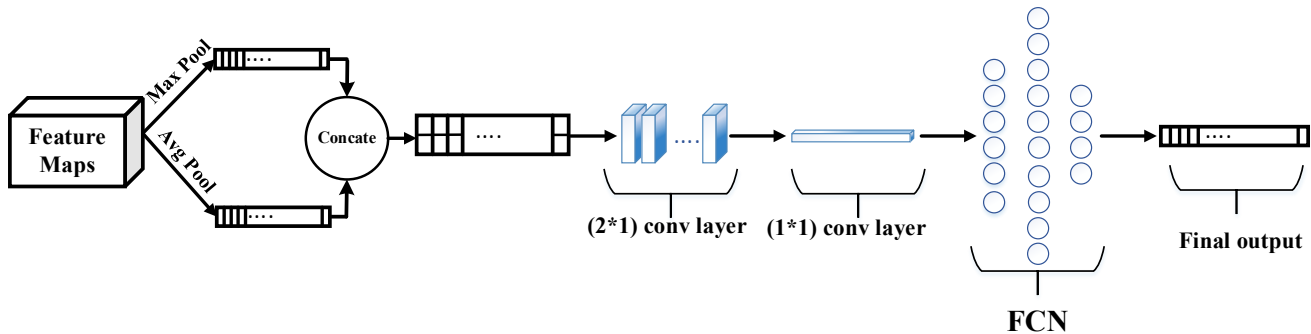

Figure 4: Channel-wise attention mechanism.

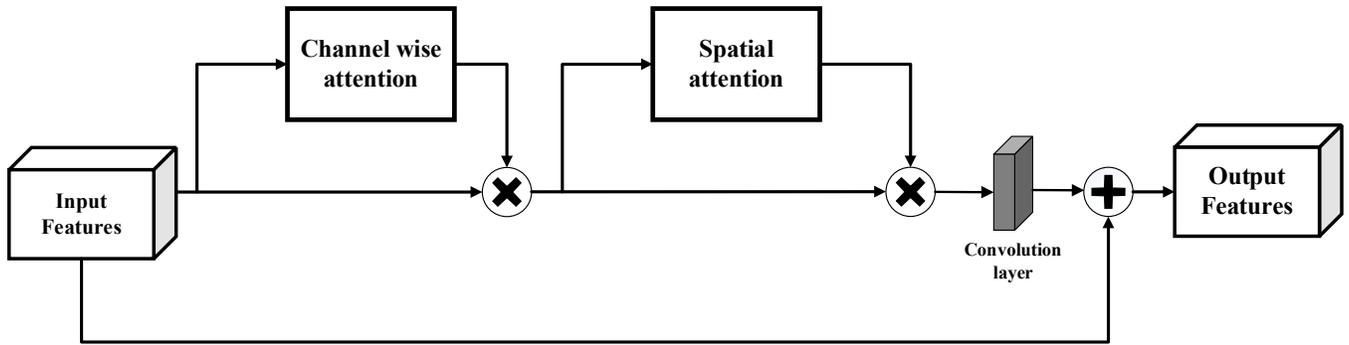

Figure 5: Fusion of attention results.

## 4. Experimental Results

There are several standard metrics that we used to evaluate our model. We use a publicly available dataset [25] for comparing our proposed method with the method in [11]. This dataset contains 20-labeled COVID-19 CT scans. The left lung, right lung, and infected regions are labeled by a radiologist and verified by another experienced radiologist. This dataset contains 20 volumes containing 3520 slices. In this dataset, 52% of slices are infected by COVID-19, and the rest are not infected. We treated this 3D dataset as a 2D dataset to reduce computational complexity. We resized each image into 256*256. For the training part, we selected 16 volumes, the same as [11], and we randomly selected 12% of the training slices for validation. For the testing phase, we also selected four volumes, the same as [11]. The results of the proposed architecture and the work in [11] are compared in Table 1.

The next experiment that we have done is using data augmentation. We combined the classical mirroring approach and deep learning-based approach for data augmentation to increase our data. On average, of each of the five folds, we generate about 600 slices totally with augmentation. Three hundred slices are generated using the classical approach and 300 slices using GANs. The results on each fold can be seen in Table 2. We trained the GAN separately for each fold and trained our model three times on each fold. The results show that using data augmentation could improve the learning process.

Since we use both classical and learning augmentation methods, in Table 3, we compare the augmentation methods on fold 1. In this experiment, we train our network once with augmented data from the classical approach and another time

Table 1. Comparing the results of our proposed method with [11] without using data augmentation.

|  |  | Fold 1 | Fold 2 | Fold 3 | Fold 4 | Fold 5 | Average±std |
|---|---|---|---|---|---|---|---|
| **Proposed Method** | **IOU** | 0.734 | 0.756 | 0.571 | 0.729 | 0.760 | **0.71±0.07** |
|  | **Dice** | 0.789 | 0.821 | 0.636 | 0.788 | 0.825 | **0.771±0.06** |
|  | **PPV** | 0.888 | 0.839 | 0.791 | 0.863 | 0.850 | **0.84±0.032** |
|  | **Sensitivity** | 0.749 | 0.819 | 0.588 | 0.763 | 0.817 | 0.747±0.084 |
|  | **Specificity** | 0.998 | 0.997 | 0.998 | 0.999 | 0.997 | 0.998±0.0006 |
| **[11]** | **IOU** | 0.719 | 0.746 | 0.594 | 0.699 | 0.752 | 0.702±0.057 |
|  | **Dice** | 0.778 | 0.812 | 0.669 | 0.762 | 0.819 | 0.768±0.053 |
|  | **PPV** | 0.831 | 0.827 | 0.764 | 0.826 | 0.852 | 0.82±0.029 |
|  | **Sensitivity** | 0.779 | 0.821 | 0.633 | 0.755 | 0.81 | **0.759±0.067** |
|  | **Specificity** | 0.998 | 0.997 | 0.997 | 0.999 | 0.998 | 0.99±0.0007 |

Table 2. Results with data augmentation.

|  | Fold 1 | Fold 2 | Fold 3 | Fold 4 | Fold 5 | Average±std |
|---|---|---|---|---|---|---|
| **IOU** | 0.759 | 0.760 | 0.636 | 0.734 | 0.771 | 0.732±0.049 |
| **Dice** | 0.812 | 0.826 | 0.703 | 0.795 | 0.836 | 0.794±0.047 |
| **PPV** | 0.884 | 0.830 | 0.789 | 0.866 | 0.859 | 0.845±0.033 |
| **Sensitivity** | 0.783 | 0.829 | 0.666 | 0.773 | 0.833 | 0.776±0.06 |
| **Specificity** | 0.999 | 0.997 | 0.997 | 0.999 | 0.998 | 0.998±0.0008 |

Table 3. Comparison between different augmentation methods on fold 1.

|  | IOU | Dice | PPV | Sensitivity | Specificity |
|---|---|---|---|---|---|
| GAN approach | **0.755** | **0.811** | 0.89 | **0.779** | 0.999 |
| Classic approach | 0.732 | 0.787 | **0.892** | 0.746 | 0.999 |

with the data from GAN. The results show the better accuracy achieved by using GAN fake data.

In our last experiment, we consider other datasets of COVID-19 CT images for infected region segmentation. There are two datasets that we use a combination of these two datasets. The first dataset [26] consists of 100 axial CT images with labels in three categories: ground-glass, consolidation, and pleural effusion. The second dataset [26] consists of 9 axial volumetric CT images which 373 slices out of the total of 829 slices have been chosen by a radiologist as positive cases. Some of the classes that are labeled in these datasets have very few slices. Although, for example, pleural class rarely occurs, like some other papers, such as [27], we consider all classes as COVID-19. After that, we combined the two datasets and randomly selected 20% of slices for testing and others for training. Our results compared to [27] can be seen in Table 4. In [27], the Focal Tversky Loss function (FTL) and deep learning (DL) are used.

## 5. Conclusion

This paper proposed a platform for segmenting the infected regions of the lung from CT images by using an image generation method for data augmentation. Our segmentation part is based on UNet, and our image generation is based on GAN. Two datasets were selected for evaluating our method. The results of both datasets showed that the proposed method produced better outputs.

In this work, we use an attention mechanism, which allows the network to focus more on the critical parts of the image. Experimental results show that this crucial information creates better results.

In addition to that, we show that data augmentation is a necessary block when the dataset is small. Among the different data augmentation approaches, the GANs can be a good choice for data augmentation. By using the data, which is generated with GANs, the model becomes more accurate and robust. The experimental results in Table 3 show that data augmentation using GAN surpasses the classical methods. It can be seen that except for PPV and specificity, the other metrics increased 2 to 3 percent.

Table 4. Comparing the results of our proposed method with [27].

|  | **IOU** | **Dice** | **PPV** | **Sensitivity** | **Specificity** |
|---|---|---|---|---|---|
| Proposed Method | 0.791 | **0.847** | 0.844 | 0.862 | 0.996 |
| [27] using FTL | - | 0.831 | - | **0.867** | **0.998** |
| [27] using DL | - | 0.815 | - | 0.767 | 0.997 |